\documentclass[]{raa}
\usepackage{graphicx,times}
\usepackage{natbib}
\usepackage{amssymb,amsmath}
\bibpunct{(}{)}{;}{a}{}{,}

\usepackage[a4paper=true,dvipdfm=true,pagebackref=true]{hyperref}
\hypersetup{pdftitle = The title of my PDF, pdfauthor = My name, pdfsubject= The subject, pdfkeywords = keyword1 keyword2 keyword3}
\hypersetup{colorlinks = true, linkcolor = green, anchorcolor = red, citecolor = blue, filecolor = red, pagecolor = red, urlcolor = red}

\begin{document}

   \title{W UMa-Type Contact Binaries in the Tidal Tails of Young Open Cluster COIN-Gaia\,25 and Mamajek\,4}

 \volnopage{ {\bf 20XX} Vol.\ {\bf X} No. {\bf XX}, 000--000}
   \setcounter{page}{1}

   \author{Xiang Bin\inst{1,2,3,4}, Liu Liang\inst{*,1,2,3,4}, Zhu Liying\inst{1,2,3,4}, Shi Xiangdong\inst{1,2,3}, Liu Nianping\inst{1,2,3}
   }

   \institute{ Yunnan Observatories, Chinese Academy of Sciences, 396 Yangfangwang, Guandu District, Kunming, P. R. China, 650216; {\it *: Corresponding Author, LiuL@ynao.ac.cn}\\
        \and
             Key Laboratory for the Structure and Evolution of Celestial Objects, Chinese Academy of Sciences, 396 Yangfangwang, Guandu District, Kunming, P. R. China, 650216\\
	    \and
             Center for Astronomical Mega-Science, Chinese Academy of Sciences, 20A Datun Road, Chaoyang District, Beijing, P. R. China, 100012\\
        \and
             University of Chinese Academy of Sciences, No.\,1 Yanqihu East Rd, Huairou District, Beijing, P. R. China, 101408\\
\vs \no
   {\small Received 20XX Month Day; accepted 20XX Month Day}
}

\abstract{Star clusters, as dynamically rich environments, are thought to be important sites for the formation of contact binaries. To investigate this, we conducted a systematic search for contact binaries within two young open clusters, COIN-Gaia\,25 and Mamajek\,4, and their associated tidal tails. From this search, we identified and confirmed two contact binary systems: ASASSN-V J064923.44$+$013758.4 in the tidal tail of COIN-Gaia\,25, and ASASSN-V J173229.06$-$613712.5 in the tidal tail of Mamajek\,4. Using TESS light curve data, we performed detailed modeling with a Markov Chain Monte Carlo (MCMC) method within the 2015 version of the Wilson-Devinney (W-D) code. The resulting photometric parameters are: $q_{\rm{ph}} = 0.316 \pm 0.013$, $i= 76.9^{+1.3}_{-0.9}$ degrees, $f=23.1^{+1.4}_{-1.9}$\,\% for ASASSN-V J064923.44$+$013758.4, and $q_{\rm{ph}} = 0.130 \pm 0.004$, $i= 68.4^{+1.4}_{-1.3}$ degrees, $f=66.3^{+5.0}_{-5.3}$\,\% for ASASSN-V J173229.06$-$613712.5. A cool-spot model located on the more massive component was successfully implemented for each system. The derived parameters classify ASASSN-V J173229.06$-$613712.5 as a deep, low-mass-ratio contact binary. Its location in the tidal tail of Mamajek\,4 constrains its age to $\leq$ 371\,Myr, supporting the view that the total lifetime of some contact binaries may be as short as a few thermal timescales. After evaluating standard formation mechanisms, we propose that ASASSN-V J173229.06$-$613712.5 likely formed via efficient orbital hardening (potentially mediated by gas dynamical friction) during the early, gas-rich phase of its host cluster's evolution. This study demonstrates the value of young clusters and their tidal tails in providing robust age constraints to explore the formation and rapid evolution of contact binaries.
}

   \authorrunning{Xiang \& Liu et al.}            
   \titlerunning{Contact Binaries in the Open Cluster Tidal Tails}  
   \maketitle

%

\section{Introduction}
Contact binaries, particularly W Ursae Majoris (W UMa) type systems, represent strong interacting binaries in which two components share a common convective envelope \citep{Lucy1968a,Lucy1968b}. They serve as important astrophysical laboratories for testing theories of binary evolution. Despite several decades of investigation, their formation and evolutionary mechanisms, particularly the lifetime of the contact phase, remain an outstanding problem in stellar astrophysics. Theoretical studies suggest that this phase typically lasts on the order of $10^8$ years, rarely exceeding $10^9$ years (e.g., \citealt{Lucy1976,Flannery1976,RobertsonEggleton1977,Stepien2011b}). In contrast, large-scale photometric surveys such as the Zwicky Transient Facility (ZTF) have revealed a striking discrepancy: the population of contact binaries exhibiting EW-type light curves significantly outnumbers that of detached binaries represented by EA-type light curves \citep{Chenetal2020}. This pronounced tension between the observed high frequency and the predicted short lifetime strongly suggests shortcomings in current evolutionary models. For instance, that the contact phase may be longer than currently estimated, or that more efficient, yet unidentified formation channels may exist.

Resolving this tension requires reliable and independent age estimates for contact binaries, which would allow direct calibration of their evolutionary timescales. Field star samples generally lack precise age determinations. Star clusters, however, provide ideal environments in which member stars share a common, well-constrained age. This makes star clusters crucial for studying the age of contact binaries. Currently, the youngest known contact binary with reliable age information is TX Cnc (e.g., \citealt{Liuetal2007,Zhangetal2009}), a member of the Praesepe Cluster (M44). The age of its host cluster is approximately 676\,Myr \citep{Kos2024}, implying that the age of this contact system itself cannot exceed this upper limit. This fact pushes the youngest reliably dated sample of contact binaries close to the theoretical upper limit of the contact phase duration ($\sim$ 1\,Gyr). It underscores the critical need to find and characterize such systems in even younger environments to better understand their rapid formation.

A promising avenue to discover such young contact binaries lies in the recently discovered tidal tail structures of open clusters. In recent years, high-precision astrometry from the Gaia satellite has led to the discovery of tidal tail structures around many open clusters (e.g., \citealt{Pangetal2022,Tarricqetal2022,Kos2024,Risbudetal2025}). These tails consist of stars stripped from their parent clusters by the Galactic tidal field, and they retain the cluster's age. Consequently, tidal tails constitute a unique, age-calibrated sample of stars well suited for studying stellar evolution under known age constraints.

Contact binaries generally possess very low orbital angular momentum. Theoretical considerations indicate that they are unlikely to have formed in their present configuration, but rather must have evolved from wider orbits through efficient angular momentum loss. Therefore, detecting contact binaries in the tidal tails of young clusters, where stellar ages are well determined, provides a direct observational constraint on the timescale for angular momentum loss from birth to contact. This timescale is critical for discriminating between different angular momentum loss mechanisms and, ultimately, for understanding the formation pathways of contact binaries.

The primary known mechanisms for angular momentum loss include gravitational wave radiation \citep{Kraft1962,Faulkner1971,Landau1975}, stellar wind mass loss, magnetic braking coupled with magnetized stellar winds \citep{Mestel1968,Skumanich1972}, common envelope evolution and ejection \citep{Paczynski1976,Iben1984}, interaction with the surrounding medium, and the Kozai-Lidov mechanism \citep{Kozai1962,Lidov1962,Fabrycky2007}.

However, gravitational wave radiation requires the stellar components to be extremely close, and common envelope evolution proceeds nearly on a dynamical timescale, rendering both mechanisms unlikely to produce stable main-sequence contact binaries. Stellar wind mass loss and interactions with the surrounding medium are predominantly significant for massive binaries. In low-mass systems, stellar winds are typically weak (e.g., \citealt{Cranmer2019,Wood2021}), and it remains unclear whether such winds are capable of sufficiently dispersing the circumstellar material to match current observational conditions. Magnetic braking and its coupling with magnetic stellar winds serve as important mechanisms for angular momentum loss in low-mass binaries evolving into contact systems, though the associated timescale is substantial approximately 10\,Gyr for solar-type stars (e.g., \citealt{vanSaders2016,Hall2021}). The Kozai-Lidov mechanism operates exclusively in triple systems. It does not directly extract angular momentum, but instead modifies the orbital architecture, thereby facilitating angular momentum loss through coupling with other processes such as tidal dissipation and gravitational wave radiation. While this form of angular momentum loss can be highly efficient, the exact amount of loss still requires accurate simulation and calculation. In this work, we therefore explore an alternative mechanism for driving angular momentum loss during contact binary formation: dynamical encounters.

In this study, we aim to identify contact binaries within young open clusters that possess detectable tidal tails. By analyzing this age-calibrated sample, we seek to directly constrain the angular momentum loss timescale necessary for the formation of contact binaries. The findings will serve as critical observational tests for evolutionary models and contribute to explaining the high observed frequency of contact binary systems.

\section{Sample Selection}
\begin{figure}
\begin{center}
	\includegraphics[angle=0,scale=0.4]{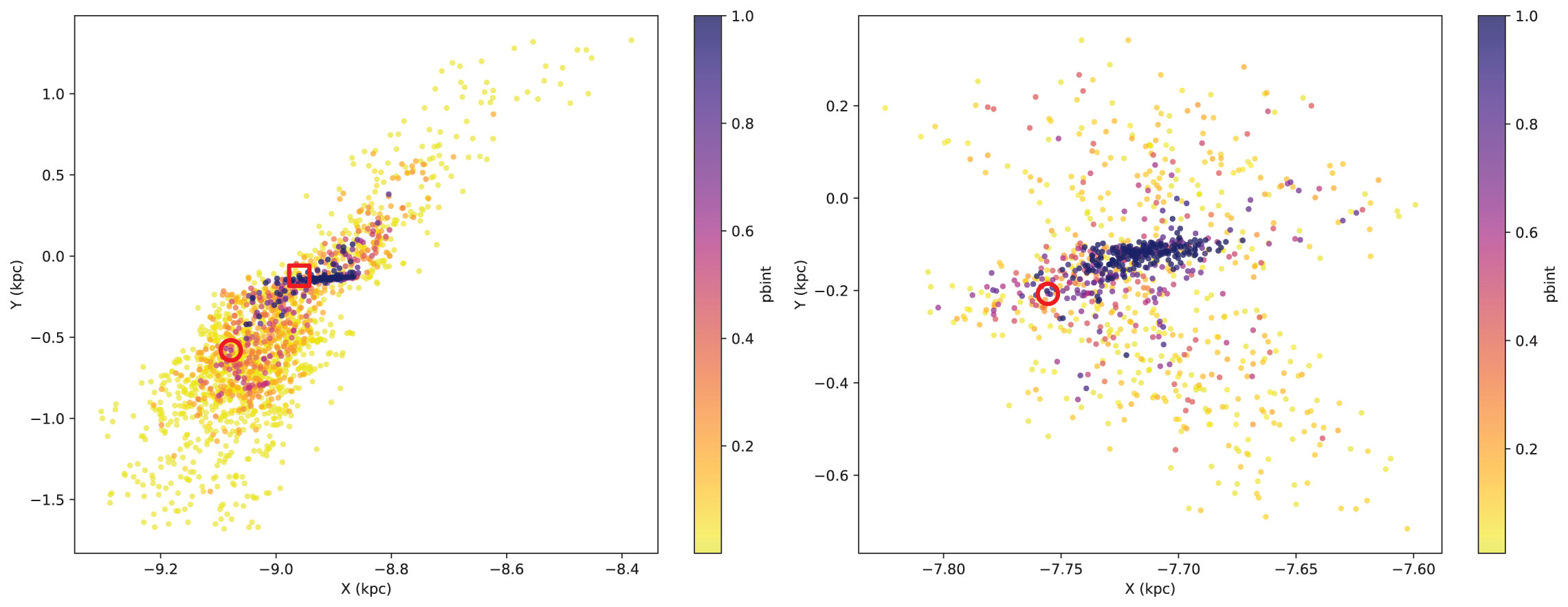}
\end{center}
 \caption{The projections of the open clusters and their tidal tail members on the plane of the Milky Way. The data are derived from \cite{Kos2024}. The left panel shows COIN-Gaia\,25, and the right panel shows Mamajek\,4.
 The red circles represent the contact binaries of ASASSN-V J064923.44$+$013758.4 and ASASSN-V J173229.06$-$613712.5, whose probability of being tidal tail members is higher than 50\,\%.
 The red square represents the contact binary candidate ZTF J060411.97$+$215302.2, with a very high membership probability but excluded by the period-color relationship.}
    \label{fig:location}
\end{figure}

The combination of Gaia's high-precision proper motion data and increasingly powerful computational resources has enabled the discovery of a growing number of stellar cluster tidal tails. A notable example is the work by \citet{Kos2024}, which identified more than one million probable tidal-tail members across 476 open clusters. By simulating the cluster dissolution process and adopting normalized membership probabilities, the author sought to reduce the threshold-dependent subjectivity often associated with likelihood-based selection methods. Their survey is complete for sources brighter than $G = 17.5$\,mag in the Gaia G-band, corresponding to an absolute $G$-band magnitude brighter than 7.5\,mag at a distance of 1\,kpc. Consequently, within this volume, their catalog should include essentially all detectable typical contact binaries. For this study, we selected two young open clusters with well-defined tidal tails, COIN-Gaia\,25 and Mamajek\,4, as the target regions. According to \cite{Kos2024}, for COIN-Gaia\,25, $N_{90\,\%}=131$, $N_{50\,\%}=229$, age = 562\,Myr, dist = 832.6\,pc, and for Mamajek\,4, $N_{90\,\%}=217$, $N_{50\,\%}=409$, age = 371\,Myr, dist = 449.6\,pc, respectively. The tidal tail member catalogs of these two clusters were cross-matched with the EW-type binary sample from the VSX catalog \citep{VSX2006} using a matching radius of 20 arcseconds. This preliminary cross-identification yielded three contact binary candidates with membership probabilities $P_{\rm{bint}} \geqslant 0.5$, as shown in Figure \ref{fig:location}.

The preliminary spatial matching only provided spatial associations for the candidates. Further screening based on the physical properties of contact binaries is necessary. We subsequently analyzed these three candidates using the period-color relation \citep{Eggen1967}. ZTF J060411.97$+$215302.2 in COIN-Gaia\,25 does not appear to be a main-sequence contact binary due to its relatively long period. The period-color relation is an important empirical relationship for contact binaries, arising from the main-sequence nature of the systems and the geometric configuration of contact. It allows contact binaries to serve as distance indicators (e.g., \citealt{MateoRucinski2017}), despite their generally low luminosities. Another key reason for choosing contact binaries as probes of cluster dynamics is their short orbital periods, which lead to a very high probability of eclipsing. Significant eclipses occur for orbital inclinations greater than about 50 degrees, and systems with an inclination angle ranging from 10 degrees to 50 degrees still exhibit light variations that are possibly similar to those of ellipsoidal variable stars. This significantly increases their detection probability, partly explaining why the number of discovered contact binaries is about four times that of other types of close binaries. Therefore, we conclude that the small number of candidates identified in the tidal tails of the two star clusters selected for this study is unlikely to be a result of observational selection bias, but rather a true reflection of the scarcity of contact binaries in these environments.

\section{Light Curve Modeling and Analysis}
\subsection{Data and Method}
Light curve modeling plays a central role in constraining the orbital parameters of contact binaries, most importantly the orbital inclination. Even in the absence of spectroscopic mass ratios, valuable physical constraints can be derived from photometric data alone. Based on numerical simulations, \cite{Liu2021} showed that for contact binaries with orbital inclinations exceeding 60 degrees, the photometric mass ratio can typically be recovered with an uncertainty better than 20\,\%. To evaluate the reliability of our solutions more rigorously, we implemented a Markov Chain Monte Carlo (MCMC; e.g., \citealt{GelfandSmith1990,Gilks1996,Gelman2013}) sampling procedure within the 2015 version of the Wilson-Devinney (W-D) code \citep{Wilson1979,Wilson1990,Wilson2008,Wilson2012,VanHammeWilson2007,Wilsonetal2010,WilsonvanHamme2014}. The MCMC approach improves the exploration of the global parameter space and provides statistically well-defined uncertainties, yielding error estimates that are mathematically more robust than those from traditional optimization methods.

\begin{figure}
\begin{center}
	\includegraphics[angle=0,scale=0.5]{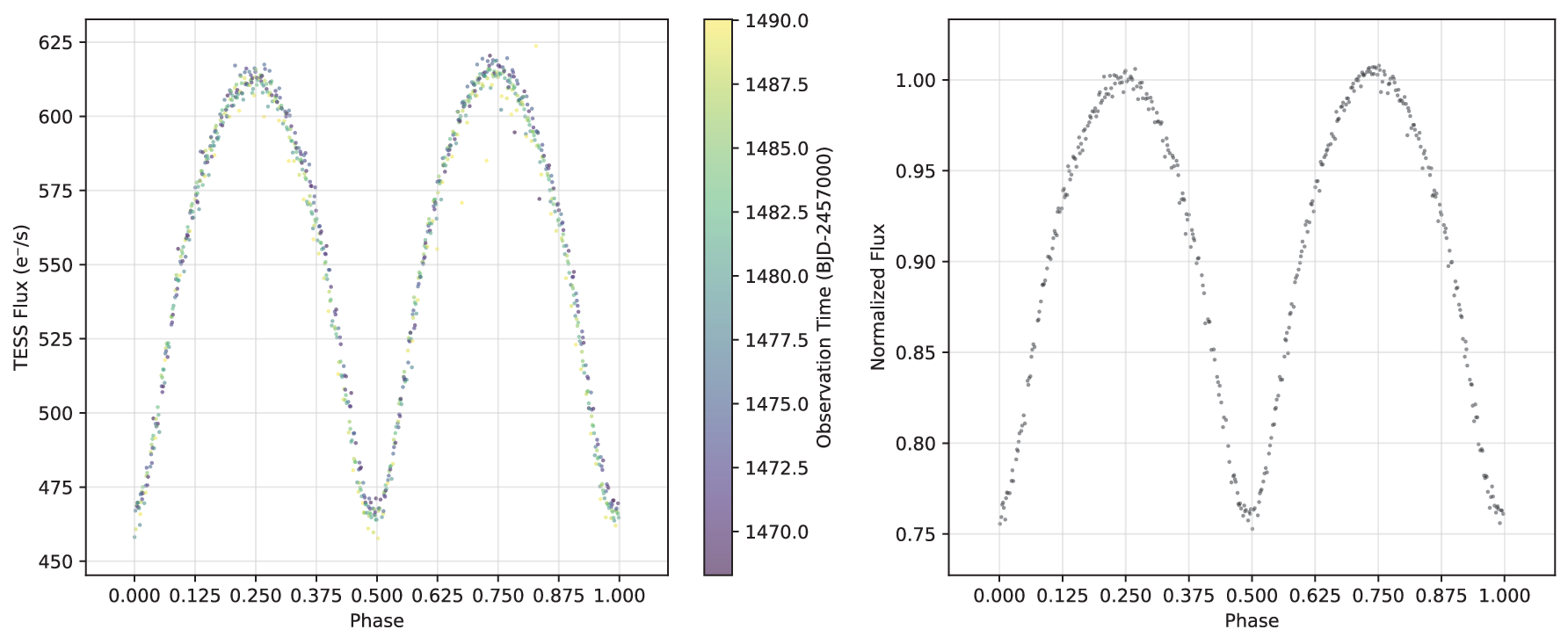}
\end{center}
 \caption{Original and normalized average TESS data for ASASSN-V J064923.44$+$013758.4, phase-folded with the period of 0.40847061\,days derived from our analysis.}
    \label{fig:lc1}
\end{figure}

\begin{figure}
\begin{center}
	\includegraphics[angle=0,scale=0.5]{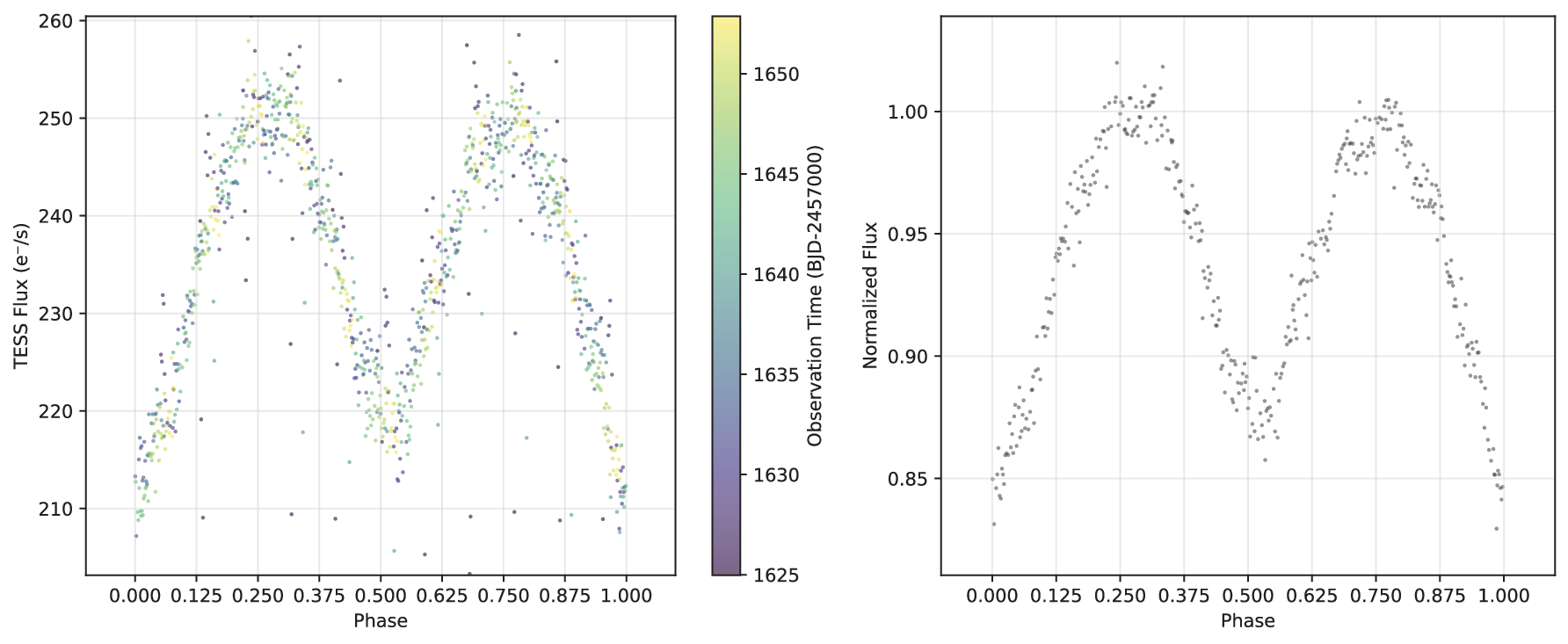}
\end{center}
 \caption{Original and normalized average TESS data for ASASSN-V J173229.06$-$613712.5, phase-folded with the period of 0.22875076\,days derived from our analysis.}
    \label{fig:lc2}
\end{figure}

The solution procedure operates as follows. The MCMC sampler explores a predefined parameter space. For each sampled parameter set, the Wilson-Devinney light curve (LC) program computes a model light curve, and the residuals relative to the observations are evaluated. The parameter set that minimizes the global residuals is adopted as the best-fit solution. The light curve data used in the fitting are from TESS \citep{Jenkins2016}, with a cadence of 30 minutes. Periods were determined via Lomb-Scargle period analysis, yielding 0.40847061\,d for ASASSN-V J064923.44$+$013758.4 and 0.22875076\,d for ASASSN-V J173229.06$-$613712.5. The phase-folded light curves are shown in Figures \ref{fig:lc1} and \ref{fig:lc2}.

\subsection{Light curve solutions and spot modeling}
\begin{figure}
\begin{center}
	\includegraphics[angle=0,scale=0.3]{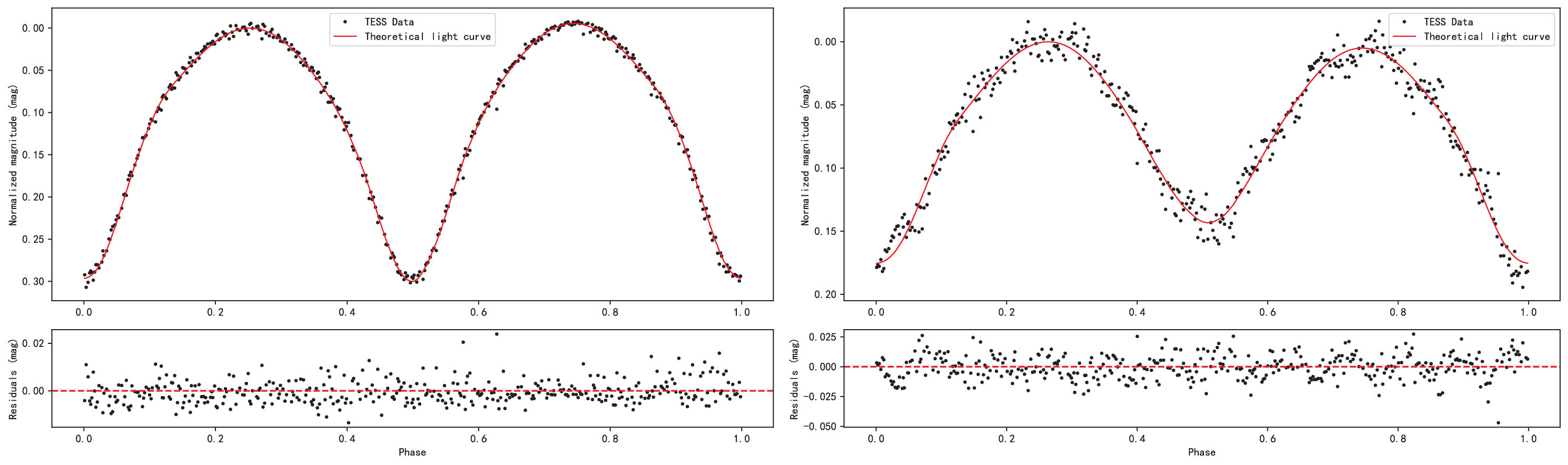}
\end{center}
 \caption{The normalized average TESS data and theoretical light curves of ASASSN-V J064923.44$+$013758.4 (left panel) and ASASSN-V J173229.06$-$613712.5 (right panel). Both systems show a cool spot on the
more massive component.}
    \label{fig:lc_fittings}
\end{figure}

\begin{figure}
\begin{center}
	\includegraphics[angle=0,scale=0.15]{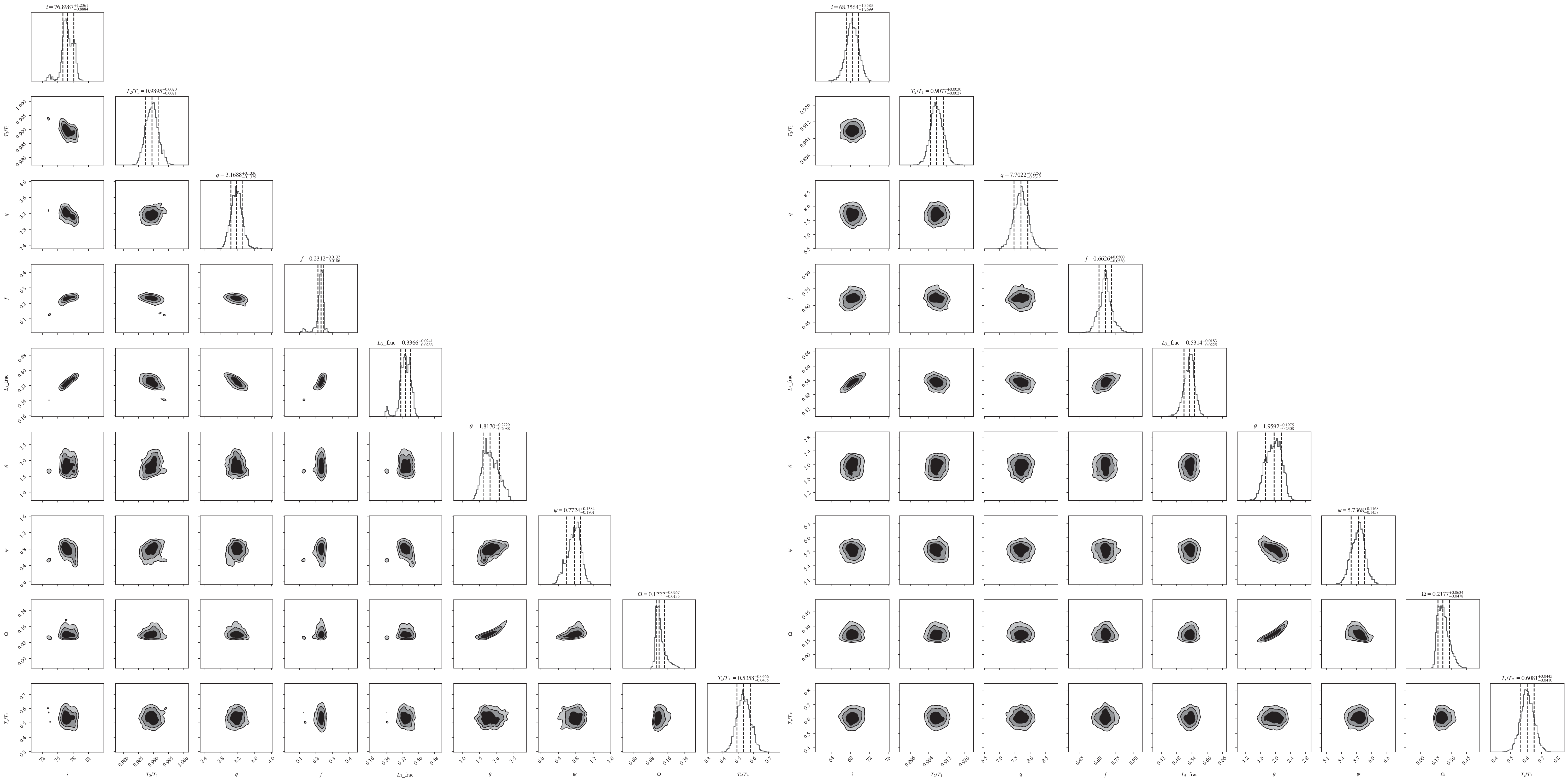}
\end{center}
 \caption{The MCMC corner plots for ASASSN-V J064923.44$+$013758.4 (left panel) and ASASSN-V J173229.06$-$613712.5 (right panel). The elliptical probability density distributions indicate positive or negative correlation between parameter pairs.}
    \label{fig:corners}
\end{figure}

During the solution, the primary temperature ($T_1$) was fixed at 5755\,K for the first target and 4596\,K for the second (Gaia DR3; \citealt{GaiaDR3Astrophysical,GaiaDR3NonSingle}). The secondary temperature ($T_2$), orbital inclination ($i$), fill-out factor ($f$), mass ratio ($q$), third light fraction ($L_{3\_\rm{frac}}$), and star spot parameters (longitude, latitude, angular radius, and temperature ratio) were treated as free parameters. The ranges of these adjusted parameters are listed in Table \ref{tab:adjusted parameters}. The gravity darkening coefficients $g_1 = g_2 = 0.32$ and the bolometric albedos $a_1 = a_2 = 0.5$ for convective common envelope were used \citep{Lucy1967}. Limb darkening coefficients were interpolated from the built-in tables of the W-D code, which depend on the local effective temperature of each star. The resulting best-fit models are overplotted on the observed light curves in Figure \ref{fig:lc_fittings}, together with the corresponding residual plots. The derived parameters are summarized in Table \ref{tab:solution parameters}. The probability density distributions of adjustable parameters are shown in Figure \ref{fig:corners}. The modeling results show third light contributions for each system. These can be accounted for by third-light contamination due to the low spatial resolution of TESS.

The asymmetry between the primary and secondary maxima in a contact binary light curve, known as the O'Connell effect \citep{O'Connell1951}, is commonly explained by star spot models. This interpretation is grounded in the properties of low-mass contact binaries: their components possess deep convective envelopes that support magnetic dynamo activity, and their extremely rapid rotation enhances magnetic phenomena. We favor the cool spot explanation over the hot spot (accretion driven) model because mass transfer in contact binaries takes place within the common convective envelope, unlike in semi-detached systems where an impacting accretion stream can produce a localized bright region. Stellar flares, which could also cause brightening, are distinguishable from hot spots by their characteristic light curve morphology and timescales. The targets studied here show no evidence of such flare-like features.

\begin{figure}
\begin{center}
	\includegraphics[angle=0,scale=0.4]{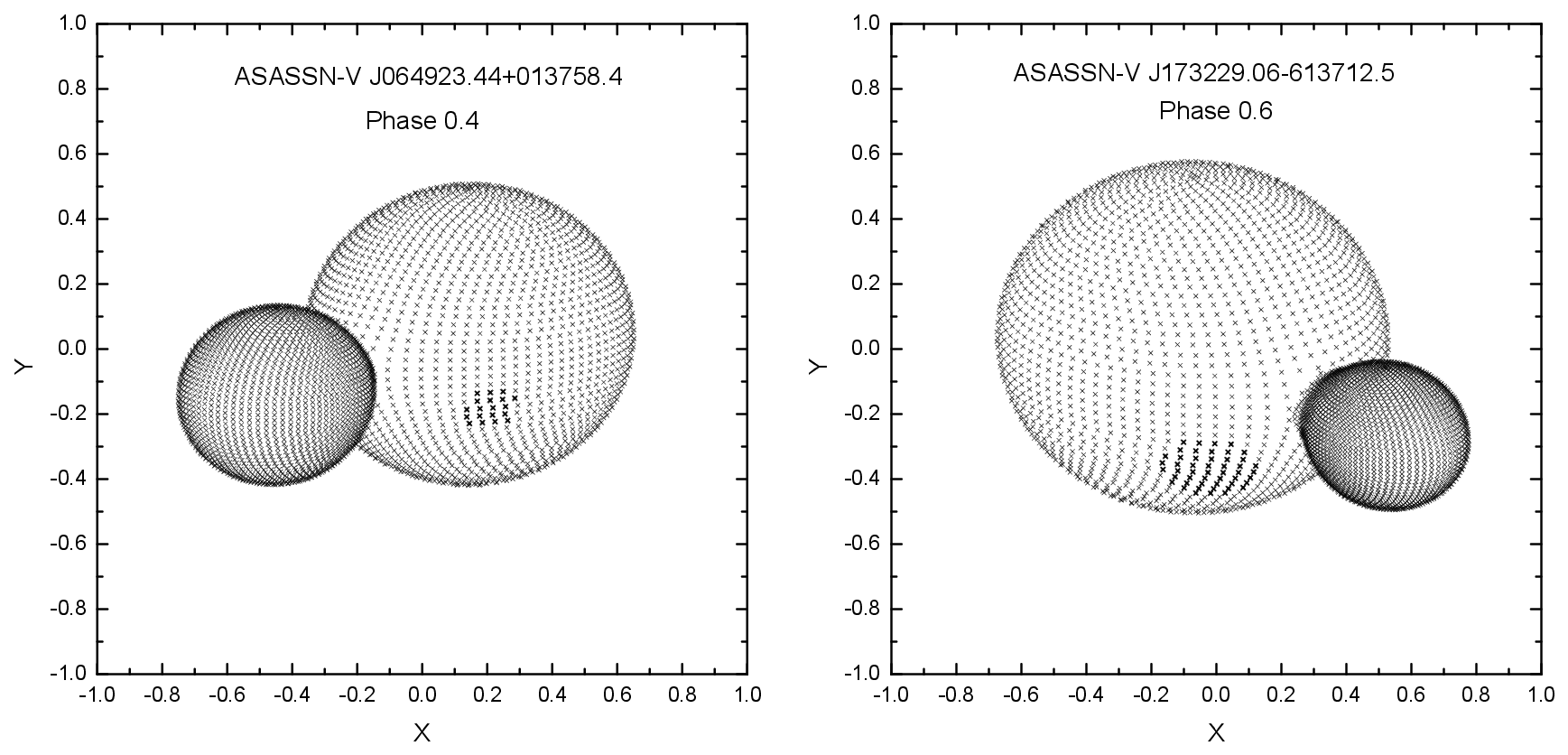}
\end{center}
 \caption{Geometrical structures for ASASSN-V J064923.44$+$013758.4 (left panel, phase 0.40) and ASASSN-V J173229.06$-$613712.5 (right panel, phase 0.60). The cool spots on their massive components are clearly visible. Both systems rotate counterclockwise when viewed from above the orbital plane.}
    \label{fig:str}
\end{figure}

We therefore applied cool spot models to both components. The results showed that a model with a spot on star 2 (the more massive component) yielded smaller residuals. The fitted spot parameters include latitude, longitude, angular radius, and temperature ratio. Their definitions can be found in the W-D documentation. Briefly, latitude is measured from the north pole (0 degree) to the south pole (180 degrees), with the direction of the star's spin angular velocity defining the north pole. Longitude is measured from the inner Lagrangian point (L1) of the Roche lobe, increasing in the direction of the star's rotation, ranging from 0 to 360 degrees. The spot angular radius is half the angle subtended by the spot radius at the center of the star, ranging from 0 to $2\pi$. A value of $2\pi$ indicates that the spot covers the entire stellar surface. The temperature ratio specifies the ratio of the local spot temperature to the local temperature that would exist without the spot. For a cool spot, the temperature ratio is less than 1. The modeling cool spots for each system are shown in Figure \ref{fig:str}.

\section{Results and Discussion}
\subsection{Derived Parameters and Physical Properties}
Based on the light curve solutions, mass ratios of $q = M_{\rm{s}}/M_{\rm{p}} = 0.316 \pm 0.013$ for ASASSN-V J064923.44$+$013758.4 and $0.130 \pm 0.004$ for ASASSN-V J173229.06$-$613712.5 were obtained. The corresponding absolute physical parameters were derived to assess the reliability of these solutions, using the same method as \cite{Liuetal2020}. We found $M_1 = 0.405\,M_{\odot}$, $M_2 = 1.285\,M_{\odot}$, $R_1 = 0.824\,R_{\odot}$, $R_2 = 1.370\,R_{\odot}$, $L_1 = 0.694\,L_{\odot}$, $L_2 = 1.867\,L_{\odot}$ for ASASSN-V J064923.44$+$013758.4, and $M_1 = 0.090\,M_{\odot}$, $M_2 = 0.692\,M_{\odot}$, $R_1 = 0.362\,R_{\odot}$, $R_2 = 0.844\,R_{\odot}$, $L_1 = 0.052\,L_{\odot}$, $L_2 = 0.178\,L_{\odot}$ for ASASSN-V J173229.06$-$613712.5. All estimated parameters are listed in Table \ref{tab:binary_parameters}. We find that these solutions are consistent with established empirical relations for main-sequence stars. ASASSN-V J173229.06$-$613712.5 is a deep, low-mass-ratio contact binary with a mass ratio of 0.130 and a fill-out factor of 66.3\,\%. The mass of its secondary component is about 0.09\,$M_{\odot}$, which is very close to the upper mass limit of brown dwarfs. In contrast, the parameters of ASASSN-V J064923.44$+$013758.4 are remarkably similar to those of TX Cnc \citep{Liuetal2007,Zhangetal2009}. Notably, the ages of their respective host clusters are also comparable. This suggests that these two contact binaries may share a common origin and have undergone similar evolutionary processes.

\subsection{The Young Age Constraint and Implications for Formation Timescales}
The contact binary ASASSN-V J173229.06$-$613712.5 resides in the tidal tail of the open cluster Mamajek\,4 (membership probability $P_{\rm{bint}}$ = 0.617), indicating that its age cannot exceed the cluster age of 371\,Myr \citep{Kos2024}. Its formation is evidently not the result of evolutionary expansion and Roche lobe overflow, because according to calculations a binary of such low total mass would require several to more than ten Gyr to evolve into contact \citep{Stepien2011b}. Magnetic braking combined with tidal dissipation likewise appears insufficient to produce a contact binary on such a short timescale. The relevant formalism for magnetic braking is described in \cite{Hurley2002}. For fully convective low-mass stars, magnetic braking is stronger, with an orbital decay timescale $\tau_{\rm{decay}} \equiv P/\dot{P} \approx (P/\rm{days})^{10/3}$ Gyr (see Appendix). When the orbital period drops below a critical threshold (typically on the order of 2\,days), the magnetic wind saturates, meaning that the associated wind torque no longer increases with faster rotation and instead approaches a nearly constant value (e.g., \citealt{Mattetal2015}). The current period of ASASSN-V J173229.06$-$613712.5 is 0.22875076\,d, placing it in the saturated magnetic braking regime. Applying the above scaling relation, if its initial period were 2 days, the corresponding decay timescale would be 2.519\,Gyr. Hence, if magnetic braking alone were responsible, the time required to shrink from a two-day orbit to the present one would far exceed the cluster age of 371\,Myr. It should be emphasized that the above magnetic braking timescales apply to fully convective stars. Most contact binaries, while possessing deep convective envelopes, are not fully convective, because full convection would likely lead to rapid merger. Consequently, the actual magnetic braking timescale driving the orbital evolution of contact binaries is probably even longer than the estimates quoted above.

\subsection{Dynamical Hardening as a Viable Formation Channel}
People naturally assume that the dense environment of a star cluster would lead to the hardening of binary stars. Could this hardening process, then, result in the formation of contact binaries? The hardening rate for binaries is given by \citep{Spitzer1987} as $d(1/a)/dt = (2\pi G \rho /\sigma)H$, with a dimensionless coefficient $H \approx 15-20$ \citep{HeggieHut2003}. According to this formula, in an open cluster of 400 members (for Mamajek\,4, $N_{90\,\%}=217$, $N_{50\,\%}=409$, \citealt{Kos2024}, with an estimated core density of $10^2\,M_{\odot}/\rm{pc}^3$), a binary like ASASSN-V J173229.06$-$613712.5 would require about 700\,Gyr to harden from an orbital separation of 20\,AU to 1.5\,$R_{\odot}$, which is far exceeding the age of the universe. Even if its core density were comparable to that of the massive young star cluster NGC\,3603 \citep{harayamaetal2008}, the same hardening formula would still yield a timescale of 12.6\,Gyr. For this binary system to complete the hardening process within 100\,Myr, the cluster core density would need to reach $10^6\,M_{\odot}/\rm{pc}^3$. From the analysis above, it is clear that dynamical encounters alone cannot produce such short-period contact binaries, which are main-sequence systems with very low angular momentum.

Star clusters were once gas-rich, and gas dynamical friction between the binary and the surrounding gas provides another effective hardening mechanism. The study by \cite{stahler2010} demonstrates that dense gas in a cluster can significantly shorten the hardening time of binaries. ASASSN-V J173229.06$-$613712.5 likely formed through gas dynamical friction. After reaching contact, it would then evolve into a deep, low-mass-ratio contact system via mass transfer.

Some might argue that a third body could also transfer angular momentum, leading the central binary to merge. However, the problem is that many simulations show such a tightly bound triple system cannot form in the first place. Disk fragmentation may be a primary mechanism for forming such low-mass hydrogen-burning binaries, but the orbital separations of binaries formed this way are typically on the order of AU \citep{Stamatellos2009}. In summary, the dense gas that once existed in star clusters likely played a crucial role in forming both short-period close binaries and contact binary systems.

\subsection{The Nature and Fate of the Deep, Low-Mass-Ratio System}
ASASSN-V J173229.06$-$613712.5 is also a deep, low-mass-ratio contact binary. \cite{Qianetal2020,YangQian2015} suggested that contact binaries with $q\leqslant0.25$, $f\geqslant50\,\%$ are deep, low-mass-ratio systems. A neural network light curve solution \citep{LiWang2025} showed that among 12201 targets, 1036 targets are with $f\geqslant50\,\%$ and 4710 targets are with $q\leqslant0.25$, while only 648 samples are deep, low-mass-ratio systems. Such systems are believed to be progenitors of luminous red novae (e.g., V1309 Sco, \citealt{Tylendaetal2011,Stepien2011a,Zhuetal2016}) then blue stragglers \citep{EggletonKiseleva-Eggleton2001,Ferreiraetal2019}.

Because of the conserved angular momentum, contact binaries should evolve towards low-mass-ratio systems. The TRO theory \citep{Lucy1976,Flannery1976,RobertsonEggleton1977} explained why there are so many shallow contact systems but did not explain why there are so many low-mass-ratio systems. Another potential mechanism \citep{Liuetal2018} could explain the phenomenon of contact binaries clustering around the low-mass-ratio region because in such a region the times of oscillation are increased, resulting in a longer timescale. Simply imposing an upper boundary on the $f$-oscillation can effectively regulate the distribution of the fill-out factor. This approach is physically plausible, since beyond a critical thickness of the common envelope the oscillatory behavior may become irreversible and cease. Nevertheless, the underlying physical mechanisms and the precise triggering conditions demand further investigation.
Regardless of which of the above processes is followed, a contact binary will gradually evolve to the deep, low-mass-ratio stage within several times the thermal timescale. The possible age 371\,Myr of ASASSN-V J173229.06$-$613712.5 is consistent with such an evolutional timescale. This contact binary may merge in a thermal timescale, producing a luminous red nova and subsequently a blue straggler.

Another question is that since the lifespan of such binary systems can only be several hundred million years, why haven't we observed a large number of their evolutionary products? The possible reason is that the luminous red novae produced during their merger did not attract much attention for a long time, while the blue stragglers formed after the merger are only easily identifiable within a stellar cluster. Given the scale of approximately hundreds of nova eruptions per year and thousands of known stellar clusters each containing at least thousands of blue stragglers, the products of contact binaries do not seem rare (e.g., \citealt{Jadhav2021}). However, more observations are still needed for verification.

\section{Summary and Conclusions}
ASASSN-V J064923.44$+$013758.4 and ASASSN-V J173229.06$-$613712.5 are contact binaries located in the tidal tails of the open clusters COIN-Gaia\,25 and Mamajek\,4, respectively. Our light curve modeling confirms their nature and yields their fundamental parameters. Notably, ASASSN-V J173229.06$-$613712.5 is a deep, low-mass-ratio contact binary. Its association with the tidal tail of Mamajek\,4 provides a robust upper age limit of 371\,Myr for its formation. This implies that the time from initial binary formation to the present deep contact configuration for systems like ASASSN-V J173229.06$-$613712.5 can be as short as a few hundred million years. The overall ages of contact binaries, and therefore the duration of the contact phase, remain active areas of research.

The short formation timescale required for ASASSN-V J173229.06$-$613712.5 presents a significant challenge to standard formation models. Our analysis demonstrates that neither isolated binary evolution (via magnetic braking) nor dynamical hardening through stellar encounters in the present-day, gas-poor cluster environment can produce such a system within 371\,Myr. This strongly suggests that alternative, more efficient formation mechanisms were operative during the early, gas-rich evolutionary stages of the star cluster. Dynamical interactions within the natal cluster environment, likely aided by gas dynamical friction, may represent a key pathway for the rapid formation of W UMa-type contact binaries. Following this rapid initial orbital decay, subsequent evolution via mass and angular momentum transfer leads systems towards the deep, low-mass-ratio configuration observed in ASASSN-V J173229.06$-$613712.5. Ultimately, such systems are prime candidates to merge as luminous red novae, subsequently enriching the population of blue stragglers.

\section*{Acknowledgments}
We thank the anonymous referee for the careful reading of the manuscript and the constructive comments, which have significantly helped to improve the quality of this work. This work is supported by the Chinese Natural Science Foundation (Nos.\,12273103 and 12573038), the Yunnan Fundamental Research Projects (grant Nos.\,202401AS070046, 202503AP140013), the International Partnership Program of Chinese Academy of Sciences (No.\,020GJHZ2023030GC) and the Yunnan Revitalization Talent Support Program.

\section*{Data Availability}
The data underlying this article are available at the Mikulski Archive for Space Telescopes (MAST) (https://mast.stsci.edu/).

\section*{Appendix: Derivation of Orbital Decay Timescale Formula}
From Kepler's third law, taking the derivative with respect to time t on both sides, we have
\begin{equation}
P^2 = \frac{4\pi^2 a^3}{GM} \quad \Rightarrow \quad \frac{\dot{a}}{a} = \frac{2}{3}\frac{\dot{P}}{P}.
\end{equation}
Do the same thing with the orbital angler momentum,
\begin{equation}
J = \mu \sqrt{GM a} \quad \Rightarrow \quad \dot{J} = \frac{\mu}{2}\sqrt{\frac{GM}{a}}\,\dot{a},
\end{equation}
where $\mu$ is the reduced mass.
According to the standard formula (\citealt{Rappaportetal1983}, formula\,36),
\begin{equation}
\dot{J}_{\rm{mb}} = -K \left(\frac{M}{M_{\odot}}\right)\left(\frac{R}{R_{\odot}}\right)^4 \left(\frac{1\,\rm{day}}{P}\right)^3.
\end{equation}
To derive the orbital decay timescale, we assume that the total angular momentum loss rate is dominated by magnetic braking ($\dot{J} = \dot{J}_{\rm{mb}}$). Using the empirical relation $R \approx M$ for low-mass fully convective stars and substituting into the orbital evolution equation $\dot{a}/a = (2/3)(\dot{P}/P)$, we obtain:
\begin{equation}
\frac{\mu}{3}\sqrt{GMa}\,\frac{\dot{P}}{P} = -K \left(\frac{M}{M_{\odot}}\right)^5\left(\frac{1\,\rm{day}}{P}\right)^3.
\end{equation}
From $a = \left(\frac{GMP^2}{4\pi^2}\right)^{1/3}$, we have $\sqrt{GMa} \propto M^{2/3}P^{1/3}$. Substituting:
\begin{equation}
\frac{\mu}{3} M^{2/3}P^{1/3} \frac{\dot{P}}{P} \propto -M^5 P^{-3}.
\end{equation}
Since $\mu \propto M$ and ignoring $M$-dependence:
\begin{equation}
\frac{\dot{P}}{P} \propto -P^{-10/3} \quad \Rightarrow \quad \tau_{\rm{decay}} \equiv \frac{P}{\dot{P}} \propto P^{10/3}.
\end{equation}
Calibration with observations gives:
\begin{equation}
\tau_{\rm{decay}} \approx \left(\frac{P}{\rm{days}}\right)^{10/3}\,\rm{Gyr}.
\end{equation}

\bibliographystyle{raa}
\bibliography{bibtex}

\begin{table}
\begin{center}
\caption{Parameter ranges for the Markov Chain Monte Carlo (MCMC) sampling.}
	\label{tab:adjusted parameters}
\begin{tiny}
\begin{tabular}{ll}
\hline
MCMC inputs                         &  Range                            \\
\hline
inclination $i$\,($^\circ$)         				& $10-90   $ \\
$T_2/T_1$                                               & $0.5-2   $ \\
$q = M_2/M_1$           				& $0.1-10  $ \\
$f=(\Omega-\Omega_{in})/(\Omega_{out}-\Omega_{in})$     & $0-1     $ \\
${L_3}\_frac=L_3/(L_1+L_2+L_3)$                        & $0-0.98  $ \\
latitude of spot $\theta$\,($^\circ$)                                    & $0-180   $ \\
longitude of spot$\psi$\,($^\circ$)                                      & $0-360  $ \\
radius of spot $\Omega\,($sr$)$                                        & $0-2\pi  $ \\
temperature ratio of spot $T_s/T_*$                                               & $0.3-3   $ \\
N\_walkers                          &                50 \\
N\_steps                             &                1000-2000 \\
\hline
\end{tabular}
\end{tiny}
\end{center}
\end{table}

\begin{table}
\begin{center}
\caption{Photometric solutions for the two contact binaries derived using the MCMC method.}
	\label{tab:solution parameters}
\begin{tiny}
\begin{tabular}{lcc}\hline
Parameters              &  ASASSN-V J064923.44$+$013758.4   & ASASSN-V J173229.06$-$613712.5 \\
\hline
$g_1=g_2$               &    0.32               								& 0.32               	         					\\
$A_1=A_2$               &    0.50               								& 0.50               	         					\\
$x_1,x_2\,(\rm TESS)$   &    $0.073, 0.087$      								& $0.359, 0.405$                     					\\
$y_1,y_2\,(\rm TESS)$   &    $0.628, 0.617$       								& $0.370, 0.343$                     					\\
$T_1$                   &    5755\,K              								& 4596\,K              	         					\\
$T_2$                   &    $5695^{\scalebox{0.5}{$+12$}}_{\scalebox{0.5}{$-13$}}$\,K 				& $4172^{\scalebox{0.5}{$+14$}}_{\scalebox{0.5}{$-12$}}$\,K 		\\
$q = M_2/M_1$           &    $3.1688  ^{\scalebox{0.5}{$+0.1336$}}_{\scalebox{0.5}{$-0.1329$}}$			& $7.7022  ^{\scalebox{0.5}{$+0.2253$}}_{\scalebox{0.5}{$-0.2312$}}$	\\
$\Omega_{in}$           &    6.8388    										& 12.4165    								\\
$\Omega_{out}$          &    6.2174    										& 11.7680    								\\
$i$\,($^\circ$)         &    $76.899^{\scalebox{0.5}{$+1.236$}}_{\scalebox{0.5}{$-0.888$}}$ 	        	& $68.356^{\scalebox{0.5}{$+1.358$}}_{\scalebox{0.5}{$-1.270$}}$ 	\\
$L_1/(L_1+L_2)   $      &    $0.270\pm0.0405$ 									& $0.2252\pm0.0389$ 							\\
$L_3/(L_1+L_2+L_3)$     &    $0.3366^{\scalebox{0.5}{$+0.0241$}}_{\scalebox{0.5}{$-0.0233$}}$ 			& $0.5314^{\scalebox{0.5}{$+0.0183$}}_{\scalebox{0.5}{$-0.0225$}}$ 	\\
$\Omega_1=\Omega_2$     &    $6.6976^{\scalebox{0.5}{$+0.0062$}}_{\scalebox{0.5}{$-0.0106$}}$			& $11.9868^{\scalebox{0.5}{$+0.0324$}}_{\scalebox{0.5}{$-0.0344$}}$	\\
$r_1(\rm{pole})$             &    $0.2747\pm0.0713$             					 		& $0.2234\pm0.0463$							\\
$r_1(\rm{side})$             &    $0.2875\pm0.0854$             					 		& $0.2354\pm0.0566$							\\
$r_1(\rm{back})$             &    $0.3285\pm0.1533$             					 		& $0.2955\pm0.0173$							\\
$r_2(\rm{pole})$             &    $0.4612\pm0.0656$             					 		& $0.5329\pm0.0364$							\\
$r_2(\rm{side})$             &    $0.4977\pm0.0902$             					 		& $0.5955\pm0.0591$							\\
$r_2(\rm{back})$             &    $0.5264\pm0.1152$             					 		& $0.6197\pm0.0712$							\\
$f$                     &    $23.1^{\scalebox{0.5}{$+1.3$}}_{\scalebox{0.5}{$-1.9$}}\,\%$			 & $66.3^{\scalebox{0.5}{$+5.0$}}_{\scalebox{0.5}{$-5.3$}}\,\%$	\\
$\theta$\,($^\circ$)    &    $104.1^{\scalebox{0.5}{$+11.0$}}_{\scalebox{0.5}{$-9.5$}}$             		& $112.3^{\scalebox{0.5}{$+11.3$}}_{\scalebox{0.5}{$-13.2$}}$           \\
$\psi$\,($^\circ$)      &    $44.3^{\scalebox{0.5}{$+7.2$}}_{\scalebox{0.5}{$-8.8$}}$               		& $328.7^{\scalebox{0.5}{$+6.7$}}_{\scalebox{0.5}{$-8.4$}}$             \\
$\Omega\,($sr$)$        &    $0.1222^{\scalebox{0.5}{$+0.0267$}}_{\scalebox{0.5}{$-0.0135$}}$        		& $0.2177^{\scalebox{0.5}{$+0.0634$}}_{\scalebox{0.5}{$-0.0478$}}$      \\
$T_s/T_*$               &    $0.5358^{\scalebox{0.5}{$+0.0466$}}_{\scalebox{0.5}{$-0.0435$}}$        		& $0.6081^{\scalebox{0.5}{$+0.0445$}}_{\scalebox{0.5}{$-0.0410$}}$      \\
rms			&  0.0049711           									&             			0.010467                        	\\
Gelman-Rubin statistic       &         1.202                                  &                      1.090     \\
\hline
\end{tabular}
\end{tiny}
\end{center}
\end{table}

\begin{table}
\begin{center}
\caption{Absolute physical parameters estimated from the light curve solutions.}
\label{tab:binary_parameters}
\begin{tabular}{lcc}
\hline
Parameter &  ASASSN-V J064923.44$+$013758.4 & ASASSN-V J173229.06$-$613712.5 \\
\hline
$T_1$ (K) & 5755 & 4596 \\
$T_2$ (K) & 5695 & 4172 \\
$M_1$ ($M_{\odot}$) & 0.405 & 0.090 \\
$M_2$ ($M_{\odot}$) & 1.285 & 0.692 \\
$R_1$ ($R_{\odot}$) & 0.824 & 0.362 \\
$R_2$ ($R_{\odot}$) & 1.370 & 0.844 \\
$L_1$ ($L_{\odot}$) & 0.694 & 0.052 \\
$L_2$ ($L_{\odot}$) & 1.867 & 0.178 \\
$A$ ($R_{\odot}$) & 2.760 & 1.450 \\
\rm{log}\,$g_1$  & 4.21 & 4.28 \\
\rm{log}\,$g_2$  & 4.27 & 4.43 \\
\hline
\end{tabular}
\end{center}
\end{table}

\end{document}